\documentclass{article} 
\usepackage{GEM_workshop_2025,times}

\usepackage{amsmath,amsfonts,bm}

\def\eqref#1{equation~\ref{#1}}

\def\1{\bm{1}}

\DeclareMathAlphabet{\mathsfit}{\encodingdefault}{\sfdefault}{m}{sl}
\SetMathAlphabet{\mathsfit}{bold}{\encodingdefault}{\sfdefault}{bx}{n}

\usepackage[bookmarks=false]{hyperref}
\usepackage{url}
\usepackage{multirow}
\usepackage{booktabs}
\usepackage{graphicx}

\title{Addressing Model Overcomplexity in Drug-Drug Interaction Prediction With Molecular Fingerprints}

\author{
Manel Gil-Sorribes$^{1}$, Alexis Molina$^{1*}$ \\
\\
$^{1}$Nostrum Biodiscovery, Barcelona, 08029, Spain \\
*\texttt{alexis.molina@nostrumbiodiscovery.com}
}

\iclrfinalcopy 
\begin{document}

\maketitle

\begin{abstract} 

Accurately predicting drug-drug interactions (DDIs) is crucial for pharmaceutical research and clinical safety. Recent deep learning models often suffer from high computational costs and limited generalization across datasets. In this study, we investigate a simpler yet effective approach using molecular representations such as Morgan fingerprints (MFPS), graph-based embeddings from graph convolutional networks (GCNs), and transformer-derived embeddings from MoLFormer integrated into a straightforward neural network. We benchmark our implementation on DrugBank DDI splits and a drug-drug affinity (DDA) dataset from the Food and Drug Administration. MFPS along with MoLFormer and GCN representations achieve competitive performance across tasks, even in the more challenging leak-proof split, highlighting the sufficiency of simple molecular representations. Moreover, we are able to identify key molecular motifs and structural patterns relevant to drug interactions via gradient-based analyses using the representations under study. Despite these results, dataset limitations such as insufficient chemical diversity, limited dataset size, and inconsistent labeling impact robust evaluation and challenge the need for more complex approaches. Our work provides a meaningful baseline and emphasizes the need for better dataset curation and progressive complexity scaling.

\end{abstract}

\section{Introduction}
Drug-drug interaction (DDI) and drug-drug affinity (DDA) predictions are essential for pharmaceutical research, impacting patient safety \citep{adverse} and drug discovery pipelines. Accurately predicting these interactions requires reliable computational methods to mitigate risks and optimize drug design \citep{computationaldd}.

Machine learning (ML) models have aimed to tackle DDI and DDA prediction beyond experimental measurement by leveraging vast chemical and biological datasets \citep{ml_ddi} \citep{ml_ddi_2}. Among these, motif-based graph learning \citep{motifsddi} and knowledge subgraph learning \citep{knowddi} introduce sophisticated architectures that have shown strong performance in DDI prediction by improving representation learning. While effective, these methods often incur high computational costs and struggle to generalize across datasets specifically designed to prevent data leakage.

A critical challenge still exists in the availability of standardized, comprehensive labeled datasets, which are essential for objectively evaluating model architectures and optimizing configurations for DDI prediction. Many current benchmarks include overlapping chemical structures between training and test sets, making it difficult to evaluate true generalization performance. Additionally, the scarcity of valuable and diverse interaction data limits the model's ability to learn meaningful patterns, as the datasets are often insufficiently large or representative to support robust training. This limitation affects the development of more efficient and interpretable approaches that can balance simplicity and predictive accuracy.

To set a meaningful baseline of low complexity, we investigate an alternative pipeline that leverages different molecular representations to capture drug interaction patterns effectively. Specifically, we utilize molecular fingerprints (MFPS) \citep{mfps} as a chemically interpretable encoding and compare them with graph-based embeddings, including pretrained GCN representations, and transformer-based embeddings from MoLFormer \citep{molformer}. These representations are evaluated within a simple neural network, providing a foundational benchmark for assessing molecular embedding effectiveness in DDI and DDA prediction. Additionally, we explore model interpretability by applying gradient-based attribution methods to trace the molecular features most influential in drug interaction predictions, allowing us to highlight chemically relevant substructures and gain insights into molecular binding mechanisms.

\section{Methodology}

\subsection{Molecule Representation}

Morgan fingerprints are a form of extended connectivity fingerprints (ECFPs), encoding molecular structures into fixed-length binary vectors that represent the presence of specific substructures. We generated Morgan fingerprints with a radius of 2 and a vector size of 2048. These embeddings are widely used due to their computational efficiency, interpretability, and effectiveness in capturing chemical similarity, making them a strong candidate for DDI  as demonstrated by \citep{mfps_top}.

GCNs represent molecules as graphs, where nodes correspond to atoms and edges define chemical bonds, with node features encoding atomic properties such as element type, hybridization state, and charge. We evaluate two configurations: a non-trained GCN, which derives embeddings without prior training and captures only raw structural information, and a pretrained GCN, trained on PCBA\_1328, a dataset of 1.6M molecules with 1,328 binary activity labels from PubChem \citep{pubchem}, allowing it to incorporate a richer chemical context.


MoLFormer is a transformer-based model tailored for molecular representation learning. It processes SMILES strings as input and captures both local and global chemical contexts. Pretrained weights from \citep{molformer} were used to extract embeddings, which were then processed through classification layers. Unlike GCNs, which operate on graph representations, MoLFormer \citep{molformer} leverages self-attention to encode molecular features from sequence data.

\subsection{Model Architecture}

To evaluate the effectiveness of different molecular representations, we designed a modular neural network that processes drug pairs and predicts their interaction outcomes. The model consists of two main components: an encoder and a classifier. The encoder is a two-layer feedforward network with batch normalization and dropout, ensuring regularization and stable training. Each drug embedding is independently transformed into a high-level representation before classification. The classifier concatenates these encoded embeddings and applies additional nonlinear transformations, with the final output computed through a softmax layer for classification or a regression layer for affinity prediction. Further architectural details and hyperparameter configurations are provided in Appendix \ref{configs_add}.

\subsection{Experimental Setup}

We evaluated two DDI classification tasks and one DDA regression task to assess the performance of different molecular embeddings within our proposed model, benchmarking against SoTA molecular graph-based and knowledge-driven approaches.

For DDI prediction, we used two dataset formulations from DrugBank \citep{drugbank}, one proposed by \citep{motifsddi} with multiple splits and another by \citep{knowddi}. 

The first data arrangement provides predefined splits, allowing a fair comparison of our model’s performance against graph-based methods. The dataset is divided into three evaluation splits:
\begin{itemize}
    \item \textbf{Unseen DDI:} The test set consists of drug pairs that do not appear in the training data, ensuring that the model predicts interactions it has never encountered.
    \item \textbf{Unseen 1 Drug:} In each test pair, one drug is absent from the training data, requiring the model to generalize interactions with previously unseen drugs.
    \item \textbf{Unseen 2 Drugs:} Both drugs in each test pair are unseen during training, representing the most challenging generalization scenario.
\end{itemize}

Additionally, we assessed performance using the DrugBank split from KnowDDI \citep{knowddi}, which introduces a structured knowledge-driven setting, enabling us to assess whether our approach remains competitive with models that incorporate explicit relational information. These two DDI benchmarks differ in data distribution and classification label sets, with full details provided in Appendixes \ref{configs_add} and \ref{dataset_volumes}.

For DDA prediction, we framed the task as a regression problem using the dataset from \citep{motifsddi}, sourced from the Food and Drug Administration (FDA). This task complements the DDI classification benchmarks by assessing how well different embeddings capture continuous interaction affinities.

\section{DDI Benchmark Results}

We assess the performance of different molecular embeddings in DDI prediction using multiple DrugBank dataset splits, comparing them against state-of-the-art (SoTA) models (Table \ref{tab:ddi_results}). Untrained GCN performed worse than other embeddings; results are in Appendix \ref{ablation_add}

\begin{table}[ht]
    \centering
    \caption{Performance of Embeddings in DDI Prediction for different splits of the DrugBank dataset. Bold values indicate the unique best result, while underscored values indicate results that are not statistically different from the best (p $>$ 0.05). GCN\_T refers to the pretrained GCN model.}
    \vspace{0.15in}
    \label{tab:ddi_results}
    \begin{tabular}{@{}lccccc@{}}
        & & & \multicolumn{3}{c}{\textbf{Our Model}} \\
        \cmidrule(lr){4-6}
        \textbf{Split} & \textbf{Metric} & \textbf{SoTA}  & \textbf{MFPS} & \textbf{GCN\_T} & \textbf{MoLF}\\
        \midrule
        \multirow{3}{*}{\shortstack{Unseen DDI}} 
        & ACC  & \textbf{97.3 \textpm 0.1}$^{1}$ & 93.2 \textpm 0.0 & 89.6 \textpm 0.0 & 92.8 \textpm 0.1\\
        & AUROC & 99.2 \textpm 0.0$^{2}$ & \textbf{99.4 \textpm 0.0} & 98.5 \textpm 0.0 & 99.3 \textpm 0.0\\
        & AUPR  & 97.9 \textpm 0.1$^{2}$ & \textbf{98.4 \textpm 0.0} & 96.5 \textpm 0.0 & 98.2 \textpm 0.0\\
        \midrule
        \multirow{3}{*}{\shortstack{Unseen 1 Drug}} 
        & ACC  & \underline{68.0 \textpm 0.7}$^{2}$ & \underline{\textbf{68.6 \textpm 0.1}} & \underline{68.6 \textpm 0.2} & 67.2 \textpm 0.2\\
        & AUROC & \textbf{88.6 \textpm 0.7}$^{3}$ & 86.3 \textpm 0.3 & 86.0 \textpm 0.2 & 85.7 \textpm 0.1\\
        & AUPR  & \underline{\textbf{73.7 \textpm 1.1}}$^{3}$ & \underline{73.1 \textpm 0.3} & 72.1 \textpm 0.2 & 71.6 \textpm 0.3\\
        \midrule
        \multirow{3}{*}{\shortstack{Unseen 2 Drugs}} 
        & ACC  & \underline{\textbf{53.6 \textpm 2.2}}$^{2}$ & \underline{52.9 \textpm 0.6} & \underline{52.9 \textpm 0.5} & 50.1 \textpm 0.7\\
        & AUROC & \textbf{78.9 \textpm 1.5}$^{2}$ & 70.9 \textpm 0.7 & 72.4 \textpm 0.4 & 70.1 \textpm 0.6\\
        & AUPR  & \underline{\textbf{53.7 \textpm 1.9}}$^{2}$ & \underline{51.5 \textpm 0.5} & \underline{51.8 \textpm 0.4} & 49.4 \textpm 0.6\\
        \bottomrule
    \end{tabular}
    \vspace{0.5em}
    \begin{minipage}{0.95\linewidth}
        \small \textit{Note:}   
        $^{1}$Results from CGIB \citep{cgib},  
        $^{2}$Results from MeTDDI \citep{motifsddi},  
        $^{3}$Results from Molormer \citep{molormer}.
    \end{minipage}
\end{table}

On the Unseen DDI split, MFPS achieves the highest Area Under the Receiver Operating Characteristic curve (AUROC) (99.4\%) and Area Under Precision Recall (AUPR) (98.4\%), closely followed by MoLFormer, outperforming all other embeddings, including the SoTA models. This indicates that fingerprint-based representations effectively capture interaction patterns when structurally similar molecules exist in both the training and test sets. The high Tanimoto similarity (0.967) in this split suggests that chemically similar drugs contribute to the elevated predictive performance.

For the Unseen 1 Drug split, MFPS and pretrained GCN embeddings achieve accuracy values (68.6\%) that are statistically indistinguishable from MeTDDI (68.0\%). Similarly, AUPR scores among these methods exhibit no significant differences, with the highest value (73.7\%) obtained by MolFormer. Despite this, MFPS and GCN embeddings show lower standard deviations, suggesting that they provide more stable predictions. This improved consistency highlights their robustness in handling partially unseen drugs.

In the Unseen 2 Drugs split, where both interacting drugs are entirely absent from training, performance drops across all methods, reflecting the challenge of predicting interactions for novel molecular pairs. While MeTDDI achieves the highest AUPR (53.7\%) and ACC (53.6\%), MFPS and GCN embeddings remain statistically comparable, with lower variance. This suggests that while absolute performance differences exist, our embeddings yield more balanced predictions with reduced variability, making them less sensitive to dataset fluctuations. Overall, while SoTA models achieve the highest individual scores in some cases, MFPS and pretrained GCN embeddings demonstrate competitive performance with lower standard deviations, offering a balanced trade-off between accuracy and stability in real-world applications. These results further reinforce the value of integrating molecular representations that balance simplicity, interpretability, and generalization ability in DDI prediction.

We further evaluated our approach by testing our simple model with different molecular embeddings on the DrugBank split proposed in \citep{knowddi}. The results were then compared with those reported for two knowledge-based models: KnowDDI \citep{knowddi} and SumGNN \citep{sumgnn}. Table \ref{tab:knowddi_results} shows that MFPS embeddings achieve the highest scores across all key metrics, outperforming knowledge-driven methods. MoLFormer embeddings also demonstrate strong performance within our model, closely trailing MFPS and significantly surpassing KnowDDI and SumGNN. These findings reinforce the effectiveness of our embedding-based approach for DDI prediction, even when compared to models that explicitly incorporate structured relational knowledge. While GCN embeddings underperform compared to MFPS and MoLFormer, they still surpass some SoTA models in certain scenarios, as detailed in Appendix \ref{ablation_add}.

\begin{table}[ht]
    \centering
    \caption{Performance of Embeddings in DrugBank (KnowDDI Benchmark).}
    \vspace{0.15in}
    \label{tab:knowddi_results}
    \begin{tabular}{@{}lccccc@{}}
        \toprule
        \textbf{Metric} & \textbf{SumGNN} & \textbf{KnowDDI} & \textbf{MFPS} & \textbf{MoLF} \\
        \midrule
        F1 & 86.88 \textpm 0.63 & 91.53 \textpm 0.24 & \textbf{95.38 \textpm 0.06} & 95.17 \textpm 0.07 \\
        ACC & 91.86 \textpm  0.23 & 93.17 \textpm 0.09 & \textbf{95.41 \textpm 0.06} & 95.21 \textpm 0.07 \\
        COHEN's $\kappa$ & 90.34 \textpm  0.28 & 91.89 \textpm 0.11 & \textbf{94.50 \textpm 0.08} & 94.26 \textpm 0.08 \\
        \bottomrule
    \end{tabular}
\end{table}

\section{Explainability of the Representations}

Identifying key molecular regions is crucial for understanding drug interactions \citep{molecular_regions}. To investigate this, we selected the drugs ritonavir and cobicistat due to their biological importance in mediating drug-drug interactions, particularly through CYP3A4/5 inhibition \citep{ritocobi_med}. We analyzed all their interaction pairs from the DrugBank dataset where they appeared as perpetrators, i.e. drugs that influence the effect of other drugs, using MFPS, pretrained GCN, and MoLFormer embeddings. These embeddings effectively capture molecular substructures relevant to drug-drug interactions, particularly in the context of cytochrome P450 inhibition. Our findings align with the proposed mechanisms of ritonavir-mediated CYP3A4 inhibition discussed in \citep{ritocobi}, reinforcing the biological relevance of the identified motifs.

\begin{table}[ht]
    \centering
    \caption{Number of DDI pairs in which each motif was identified as relevant for predictions with MFPS and MoLFormer embeddings. Extended results in Appendix \ref{motif_add}.}
    \vspace{0.15in}
    \label{tab:sulfur_motifs_combined}
    \begin{tabular}{@{}lcccc@{}}
         & \multicolumn{2}{c}{\textbf{Cobicistat} (469 pairs)} & \multicolumn{2}{c}{\textbf{Ritonavir} (337 pairs)} \\
        \cmidrule(lr){2-3} \cmidrule(lr){4-5}
        \textbf{Motif} & \textbf{MFPS} & \textbf{MoLF} & \textbf{MFPS} & \textbf{MoLF} \\
        \midrule
        Thiazole core (ccsc(C)n) & 292 & - & 259 & - \\
        Substituted thiazole (cc(C)nc(C)s) & 276 & - & 206 & - \\
        cnc(cs)CN & 100 & - & 110 &\\
        Cc(n)s & 1 & 157 & - & 27\\
        \bottomrule
    \end{tabular}
\end{table}

We applied gradient-based attribution to identify molecular substructures contributing to interaction predictions \citep{gradient-based}. For MFPS, this method mapped fingerprint bits to their corresponding substructures, highlighting key regions. For MoLFormer, we used attention scores to pinpoint critical atoms and bonds. For GCN embeddings, we traced node and edge importance within molecular graphs to understand structural contributions. 

This analysis revealed several recurring sulfur-containing motifs in the interaction pairs involving ritonavir and cobicistat, both known for their CYP3A4/5 inhibition activity. In our dataset, ritonavir appeared as a perpetrator in 337 DDI pairs, while cobicistat did so in 469 pairs. Table \ref{tab:sulfur_motifs_combined} shows the frequency with which these motifs were detected across the DDI pairs. MFPS captured thiazole cores and sulfur-nitrogen motifs essential for CYP3A4 inhibition in ritonavir and cobicistat. Notably, MFPS identified sulfur-nitrogen motifs (cnc(cs)CN) in both molecules, linked to its role as a mechanism-based inhibitor. MoLFormer emphasized a thiazole variant (Cc(n)s), indicating an alternative recognition pattern. Our results are consistent with findings in \citet{motifsddi}. Ritonavir’s CYP3A4 inhibition remains actively studied, with mechanisms including tight heme binding and covalent modification \citep{ritocobi}. Sulfur motifs play a key role through Metabolic-Intermediate Complex formation, heme ligation, or covalent modification. The identified motifs align with known pathways, highlighting the relevance of these substructures in DDI.

\section{DTA Benchmark Results}

Accurate affinity prediction is essential for understanding DDI at a quantitative level. We evaluated different molecular representations on this task to assess their effectiveness in capturing interaction strength, using a dataset sourced from the FDA. The variable to predict is $\log_2(\text{AUC FC})$, where AUC FC represents the fold change of the area under the plasma time–concentration curve. MFPS achieved the highest Pearson correlation (PCC) (0.831) and the lowest Root Mean Square Error (RMSE) (0.838) (visualization in Appendix \ref{dda_mfps_plot}), demonstrating its reliability in modeling DDI. As shown in Table \ref{tab:dta_results}, MFPS outperforms MeTDDI, which reports a lower PCC and a higher RMSE.

Other DL-based models, such as SA-DDI \citep{saddi}, Molormer \citep{molormer} and MolTrans \citep{moltrans}, show further declines in predictive accuracy. These findings underscore the robustness of our model using MFPS in affinity prediction, demonstrating that a chemically grounded fingerprint-based approach can surpass complex transformer-based models in capturing meaningful molecular interactions. Results for GCN, pretrained GCN and MoLFormer embeddings are reported in Appendix \ref{ablation_add}, where they exhibited lower performance across all evaluated metrics, further emphasizing the effectiveness of fingerprint-based representations.

\begin{table}[h!]
    \centering
    \caption{Performance on affinity prediction in the FDA benchmark.}
    \vspace{0.15in}
    \label{tab:dta_results}
    \begin{tabular}{@{}lcc@{}}
        \toprule
        \textbf{Embedding} & \textbf{RMSE} & \textbf{PCC} \\
        \midrule
        Ours (MFPS) & \textbf{0.838} & \textbf{0.831} \\
        MeTDDI & 0.912 & 0.725 \\
        SA-DDI & 0.961 & 0.724 \\
        Molormer & 1.013 & 0.721 \\
        MolTrans & 1.032 & 0.690 \\
        \bottomrule
    \end{tabular}
\end{table}

\section{Conclusions}
We proposed a simple yet effective model with mildly complex representations for drug-drug interaction (DDI) and drug-drug affinity (DDA) prediction. Despite its simplicity, our approach achieves competitive results, with MFPS and pretrained GCN embeddings often matching or surpassing state-of-the-art models that rely on significantly more complex architectures. This demonstrates that high performance can be achieved without excessive model complexity, offering a more efficient and interpretable alternative.

However, evaluating these models remains challenging due to dataset limitations. While some benchmarks emphasize leak-proof splits, they often lack sufficient chemical diversity, dataset scale, and consistent labeling, making it difficult to accurately assess generalization and determine the optimal model complexity for the task. Our study highlights the need for both rigorous dataset curation and methodical complexity scaling. Future work should focus on expanding dataset coverage, ensuring high-quality interaction labels, and incorporating advanced techniques to progressively increase model complexity while maintaining interpretability and efficiency.


\bibliography{iclr2025_conference}

\begin{thebibliography}{20}
\providecommand{\natexlab}[1]{#1}
\providecommand{\url}[1]{\texttt{#1}}
\expandafter\ifx\csname urlstyle\endcsname\relax
  \providecommand{\doi}[1]{doi: #1}\else
  \providecommand{\doi}{doi: \begingroup \urlstyle{rm}\Url}\fi

\bibitem[Bissantz et~al.(2010)Bissantz, Kuhn, and Stahl]{molecular_regions}
Caterina Bissantz, Bernd Kuhn, and Martin Stahl.
\newblock A medicinal chemist's guide to molecular interactions.
\newblock \emph{Journal of Medicinal Chemistry}, 53\penalty0 (14):\penalty0 5061--5084, Jul 2010.

\bibitem[Huang et~al.(2021)Huang, Xiao, Glass, and Sun]{moltrans}
Kexin Huang, Cao Xiao, Lucas~M Glass, and Jimeng Sun.
\newblock {MolTrans}: molecular interaction transformer for drug--target interaction prediction.
\newblock \emph{Bioinformatics}, 37\penalty0 (6):\penalty0 830--836, 2021.

\bibitem[Jiang et~al.(2022)Jiang, Lin, Ren, Fang, Liu, Tan, Lv, and Zhang]{adverse}
Huaqiao Jiang, Yanhua Lin, Weifang Ren, Zhonghong Fang, Yujuan Liu, Xiaofang Tan, Xiaoqun Lv, and Ning Zhang.
\newblock Adverse drug reactions and correlations with drug--drug interactions: A retrospective study of reports from 2011 to 2020.
\newblock \emph{Frontiers in Pharmacology}, 13:\penalty0 923939, 2022.

\bibitem[Jim{\'e}nez-Luna et~al.(2020)Jim{\'e}nez-Luna, Grisoni, and Schneider]{gradient-based}
Jos{\'e} Jim{\'e}nez-Luna, Francesca Grisoni, and Gisbert Schneider.
\newblock Drug discovery with explainable artificial intelligence.
\newblock \emph{Nature Machine Intelligence}, 2\penalty0 (10):\penalty0 573--584, Oct 2020.
\newblock ISSN 2522-5839.
\newblock \doi{10.1038/s42256-020-00236-4}.

\bibitem[Kim et~al.(2023)Kim, Chen, Cheng, Gindulyte, He, He, Li, Shoemaker, Thiessen, Yu, et~al.]{pubchem}
Sunghwan Kim, Jie Chen, Tiejun Cheng, Asta Gindulyte, Jia He, Siqian He, Qingliang Li, Benjamin~A Shoemaker, Paul~A Thiessen, Bo~Yu, et~al.
\newblock {PubChem} 2023 update.
\newblock \emph{Nucleic Acids Research}, 51\penalty0 (D1):\penalty0 D1373--D1380, 2023.

\bibitem[Lee et~al.(2023)Lee, Hyun, Na, Kim, Lee, and Park]{cgib}
Namkyeong Lee, Dongmin Hyun, Gyoung~S Na, Sungwon Kim, Junseok Lee, and Chanyoung Park.
\newblock Conditional graph information bottleneck for molecular relational learning.
\newblock In \emph{International Conference on Machine Learning}, pp.\  18852--18871. PMLR, 2023.

\bibitem[Lin et~al.(2022)Lin, Wang, Zhang, Chu, Liu, Fang, Jiang, Wang, Zhao, Xiong, et~al.]{saddi}
Shenggeng Lin, Yanjing Wang, Lingfeng Zhang, Yanyi Chu, Yatong Liu, Yitian Fang, Mingming Jiang, Qiankun Wang, Bowen Zhao, Yi~Xiong, et~al.
\newblock {MDF-SA-DDI}: predicting drug--drug interaction events based on multi-source drug fusion, multi-source feature fusion and transformer self-attention mechanism.
\newblock \emph{Briefings in Bioinformatics}, 23\penalty0 (1):\penalty0 bbab421, 2022.

\bibitem[Long et~al.(2022)Long, Pan, Zhang, Song, Kondor, and Rzhetsky]{mfps_top}
Yanan Long, Horace Pan, Chao Zhang, Hy~Truong Song, Risi Kondor, and Andrey Rzhetsky.
\newblock Molecular fingerprints are a simple yet effective solution to the drug--drug interaction problem.
\newblock \emph{Drugs}, 500:\penalty0 1--7, 2022.

\bibitem[Loos et~al.(2022)Loos, Beijnen, and Schinkel]{ritocobi}
Nancy~HC Loos, Jos~H Beijnen, and Alfred~H Schinkel.
\newblock The mechanism-based inactivation of cyp3a4 by ritonavir: what mechanism?
\newblock \emph{International Journal of Molecular Sciences}, 23\penalty0 (17):\penalty0 9866, 2022.

\bibitem[Marzolini et~al.(2016)Marzolini, Gibbons, Khoo, and Back]{ritocobi_med}
Catia Marzolini, Sara Gibbons, Saye Khoo, and David Back.
\newblock Cobicistat versus ritonavir boosting and differences in the drug-drug interaction profiles with co-medications.
\newblock \emph{J. Antimicrob. Chemother.}, 71\penalty0 (7):\penalty0 1755--1758, jul 2016.

\bibitem[Ou-Yang et~al.(2012)Ou-Yang, Lu, Kong, Liang, Luo, and Jiang]{computationaldd}
Si-sheng Ou-Yang, Jun-yan Lu, Xiang-qian Kong, Zhong-jie Liang, Cheng Luo, and Hualiang Jiang.
\newblock Computational drug discovery.
\newblock \emph{Acta Pharmacologica Sinica}, 33\penalty0 (9):\penalty0 1131--1140, 2012.

\bibitem[Rogers \& Hahn(2010)Rogers and Hahn]{mfps}
David Rogers and Mathew Hahn.
\newblock Extended-connectivity fingerprints.
\newblock \emph{Journal of Chemical Information and Modeling}, 50\penalty0 (5):\penalty0 742--754, 2010.

\bibitem[Ryu et~al.(2018)Ryu, Kim, and Lee]{ml_ddi_2}
Jae~Yong Ryu, Hyun~Uk Kim, and Sang~Yup Lee.
\newblock Deep learning improves prediction of drug--drug and drug--food interactions.
\newblock \emph{Proceedings of the National Academy of Sciences}, 115\penalty0 (18):\penalty0 E4304--E4311, 2018.

\bibitem[Wang et~al.(2024)Wang, Yang, and Yao]{knowddi}
Yaqing Wang, Zaifei Yang, and Quanming Yao.
\newblock Accurate and interpretable drug-drug interaction prediction enabled by knowledge subgraph learning.
\newblock \emph{Communications Medicine}, 4\penalty0 (1):\penalty0 59, Mar 2024.

\bibitem[Wishart et~al.(2006)Wishart, Knox, Guo, Shrivastava, Hassanali, Stothard, Chang, and Woolsey]{drugbank}
David~S Wishart, Craig Knox, An~Chi Guo, Savita Shrivastava, Murtaza Hassanali, Paul Stothard, Zhan Chang, and Jennifer Woolsey.
\newblock {DrugBank}: a comprehensive resource for in silico drug discovery and exploration.
\newblock \emph{Nucleic Acids Research}, 34\penalty0 (suppl\_1):\penalty0 D668--D672, 2006.

\bibitem[Wu et~al.(2023)Wu, Radev, and Li]{molformer}
Fang Wu, Dragomir Radev, and Stan~Z Li.
\newblock Molformer: Motif-based transformer on {3D} heterogeneous molecular graphs.
\newblock In \emph{Proceedings of the AAAI Conference on Artificial Intelligence}, volume~37, pp.\  5312--5320, 2023.

\bibitem[Yang et~al.(2022)Yang, Zhong, Lv, and Chen]{ml_ddi}
Ziduo Yang, Weihe Zhong, Qiujie Lv, and Calvin Yu-Chian Chen.
\newblock Learning size-adaptive molecular substructures for explainable drug--drug interaction prediction by substructure-aware graph neural network.
\newblock \emph{Chemical Science}, 13\penalty0 (29):\penalty0 8693--8703, 2022.

\bibitem[Yu et~al.(2021)Yu, Huang, Zhang, Glass, Sun, and Xiao]{sumgnn}
Yue Yu, Kexin Huang, Chao Zhang, Lucas~M Glass, Jimeng Sun, and Cao Xiao.
\newblock {SumGNN}: multi-typed drug interaction prediction via efficient knowledge graph summarization.
\newblock \emph{Bioinformatics}, 37\penalty0 (18):\penalty0 2988--2995, 2021.

\bibitem[Zhang et~al.(2022)Zhang, Wang, Meng, Wang, Zhang, Rodriguez-Paton, Wang, and Wang]{molormer}
Xudong Zhang, Gan Wang, Xiangyu Meng, Shuang Wang, Ying Zhang, Alfonso Rodriguez-Paton, Jianmin Wang, and Xun Wang.
\newblock Molormer: a lightweight self-attention-based method focused on spatial structure of molecular graph for drug--drug interactions prediction.
\newblock \emph{Briefings in Bioinformatics}, 23\penalty0 (5):\penalty0 bbac296, 2022.

\bibitem[Zhong et~al.(2024)Zhong, Li, Yang, Zheng, Yu, Zhang, Luo, Wang, and Weng]{motifsddi}
Yi~Zhong, Gaozheng Li, Ji~Yang, Houbing Zheng, Yongqiang Yu, Jiheng Zhang, Heng Luo, Biao Wang, and Zuquan Weng.
\newblock Learning motif-based graphs for drug-drug interaction prediction via local-global self-attention.
\newblock \emph{Nature Machine Intelligence}, 6:\penalty0 1094--1105, 2024.

\end{thebibliography}
\bibliographystyle{iclr2025_conference}

\newpage
\appendix

\section{Model Architectures and Hyperparameter Configurations}\label{configs_add}

To evaluate embeddings of drug pairs, we designed a modular neural network for drug-drug interaction (DDI) prediction. The architecture consists of:

\begin{itemize}
    \item \textbf{Encoder:} A two-layer feedforward network incorporating batch normalization and dropout for regularization. Each drug embedding is independently processed to extract high-level features.
    \item \textbf{Classifier:} A fully connected network that concatenates the encoded embeddings, applies nonlinear transformations, and outputs interaction probabilities via a softmax layer.
\end{itemize}

The model supports embeddings with different dimensionalities (emb. dim.) depending on their origin:

\begin{itemize}
    \item \textbf{Morgan fingerprints (MFPS):} Binary vectors with a fixed dimensionality of 2048, providing interpretable and chemically intuitive features.
    \item \textbf{MoLFormer embeddings:} Dense representations of dimension 768, derived from a transformer-based model trained on molecular sequences.
    \item \textbf{GCN embeddings:} Compact embeddings with a dimensionality of 64, extracted from a pretrained graph convolutional network that captures structural relationships between atoms and bonds.
\end{itemize}

\begin{figure}[ht]
    \centering
    \includegraphics[width=0.8\textwidth]{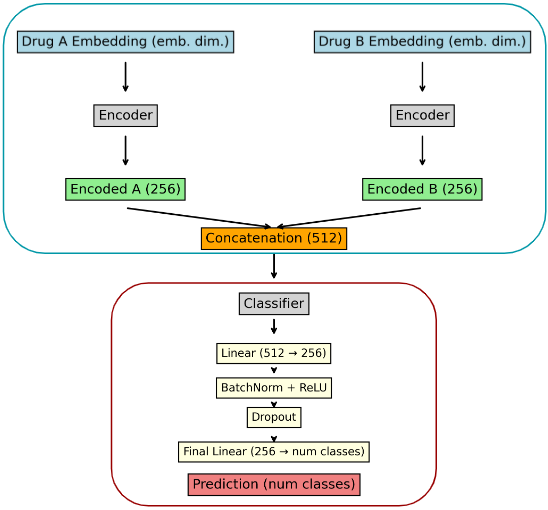}
    \caption{Neural network architecture for DDI prediction, consisting of an encoder for feature extraction and a classifier for interaction classification.}
    \label{fig:architecutre}
\end{figure}

\begin{table}[ht]
    \centering
    \caption{Hyperparameters used across different settings.}
    \vspace{0.15in}
    \label{tab:hyperparameters}
    \begin{tabular}{@{}lccc@{}}
        \toprule
        \textbf{Hyperparameter} & \textbf{DDI Benchmark 1} & \textbf{DDI Benchmark 2} & \textbf{DDA Regression} \\
        \midrule
        Embedding Dimension & emb. dim. & emb. dim. & emb. dim. \\
        Hidden Dimension & 512 & 512 & 1024 \\
        Output Dimension & 256 & 256 & 512 \\
        Number of Classes & 4 & 85 & 1 (Regression) \\
        Dropout Rate & 0.3 & 0.3 & 0.1 \\
        Learning Rate & $1 \times 10^{-4}$ & $1 \times 10^{-4}$ & $1 \times 10^{-5}$ \\
        Batch Size & 512 & 512 & 512 \\
        Optimizer & Adam & Adam & Adam \\
        Loss Function & CrossEntropyLoss & CrossEntropyLoss & MSELoss \\
        Number of Epochs & 100 & 100 & 200 \\
        \bottomrule
    \end{tabular}
\end{table}

The model maintains a consistent architecture across tasks while adapting the embedding dimension, number of classes, and learning rate for each benchmark. Regularization techniques, such as dropout and batch normalization, are applied to enhance generalization across datasets.

\begin{table}[ht]
    \centering
    \caption{Comparison of Our Model with More Complex Drug-Drug Interaction (DDI) Models.}
    \vspace{0.15in}
    \label{tab:model_comparison}
    \renewcommand{\arraystretch}{1.2}
    \begin{tabular}{@{}lcc@{}}
        \toprule
        \textbf{Model} & \textbf{Complexity} & \textbf{Key Features} \\
        \midrule
        \textbf{Simple Model} (MFPS + GCN) & $\downarrow$  (Shallow NN, no attention) & Fast, interpretable, low resource usage \\
        MeTDDI (Motif-based GNN) & $\uparrow$  (Self-attention, graph reasoning) & Captures local/global patterns in graphs \\
        KnowDDI (KG + GNN) & $\Uparrow$ (Subgraph learning, pretraining) & Uses relational data, high scalability \\
        \bottomrule
    \end{tabular}
\end{table}

\newpage

\section{Dataset Sizes Across Benchmarks} \label{dataset_volumes}

\begin{figure}[ht]
    \centering
    \includegraphics[width=\textwidth]{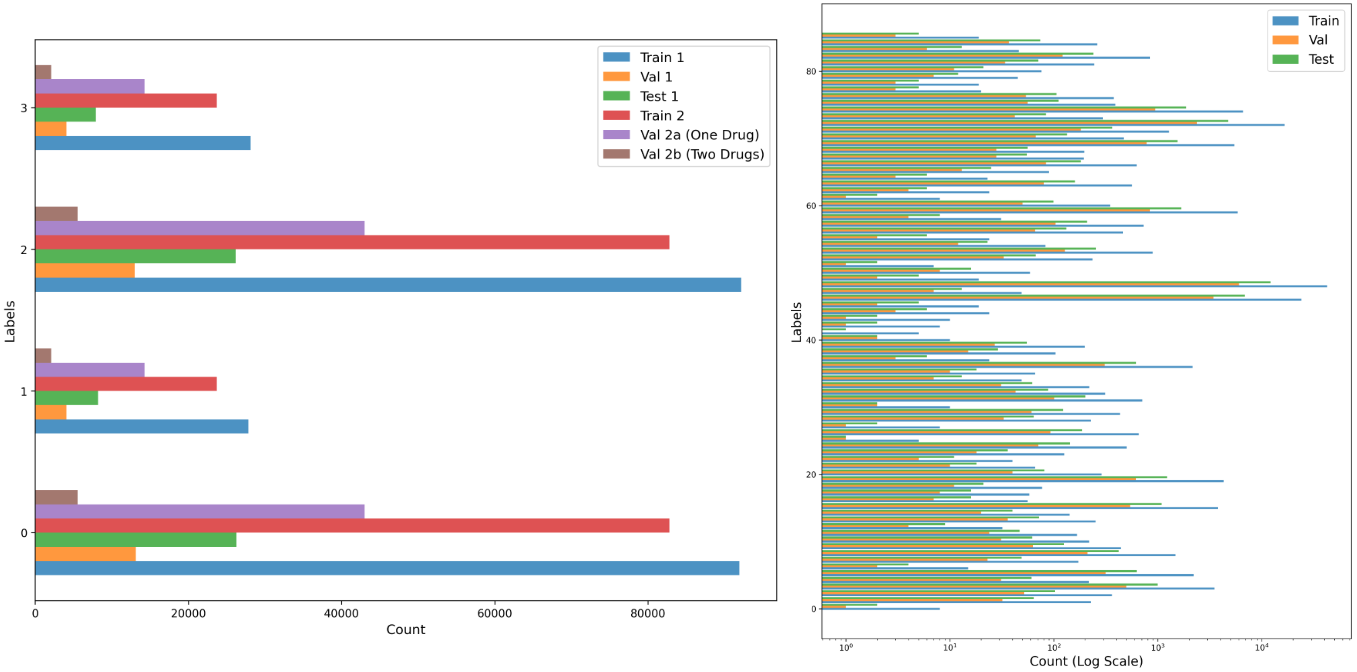}
    \caption{
        Label distribution across datasets. 
        \textbf{Left:} Distribution of the first proposed dataset splits, with labels defined according to \citep{motifsddi}. 
        \textbf{Right:} Distribution in the split proposed by \citep{knowddi}, using its respective label definitions.
    }
    \label{fig:label_distribution}
\end{figure}

\begin{table}[ht]
    \centering
    \caption{Dataset volumes for DDI and DDA benchmarks across train, validation, and test splits. All DDI datasets are derived from DrugBank, while the DDA dataset is sourced from FDA data. The numbers represent drug pairs.}
    \vspace{0.15in}
    \label{tab:dataset_volumes}
    \begin{tabular}{@{}lcccccc@{}}
        \toprule
        \textbf{Task} & & \textbf{Dataset} & & \textbf{Train} & \textbf{Validation} & \textbf{Test} \\
        \midrule
        \multirow{3}{*}{\shortstack{\textbf{DDI} \\ \textbf{(DrugBank)}}}
        & & Unseen DDI & & 240,124 & 34,304 & 68,608 \\ 
        & & One-Drug Unseen & & 213,104 & - & 114,614 \\ 
        & & Two-Drugs Unseen & & 213,104 & - & 15,318 \\ 
        \cmidrule(lr){3-3}
        & & KnowDDI (Knowledge Graph) & & 69,718 & 10,041 & 19,755 \\ 
        \midrule
        \textbf{DDA (FDA)} & & Affinity Prediction & & 3,189 & 798 & 47 \\ 
        \bottomrule
    \end{tabular}
\end{table}

\section{Further Benchmarks on Embedding Effectiveness}\label{ablation_add}

We present a full comparison of all tested embeddings, evaluating their performance in both DDI and DDA benchmarks. All our results are averaged over five independent training runs, with mean and standard deviation reported. Results for other models are taken from their respective studies.

\subsection{Benchmark vs Graph-Based Models}

\begin{table}[ht]
    \centering
    \caption{Ablation Study: Performance of Different Embeddings in DDI Prediction for different splits of the DrugBank dataset proposed by \citep{motifsddi}. GCN\_T refers to the pretrained GCN model.}
    \vspace{0.15in}
    \label{tab:ddi_ablation_results}
    \begin{tabular}{@{}lcccccc@{}}
        & & & &  \multicolumn{2}{c}{\textbf{Our Model}} \\
        \cmidrule(lr){4-7}
        \textbf{Split} & \textbf{Metric} & \textbf{SoTA} & \textbf{MFPS} & \textbf{GCN\_T} & \textbf{GCN} & \textbf{MoLF}\\
        \midrule
        \multirow{3}{*}{\shortstack{Unseen\\DDI}} & ACC  & 97.3 \textpm 0.1$^1$ & 93.2 \textpm 0.0 & 89.6 \textpm 0.0 & 88.9 \textpm 0.0 & 92.8 \textpm 0.1\\
                                     & AUROC & 99.2 \textpm 0.0$^2$ & 99.4 \textpm 0.0 & 98.5 \textpm 0.0 & 98.4 \textpm 0.0 & 99.3 \textpm 0.0\\
                                     & AUPR  & 97.9 \textpm 0.1$^2$ & 98.4 \textpm 0.0 & 96.5 \textpm 0.0 & 96.2 \textpm 0.0 & 98.2 \textpm 0.0\\
        \midrule
        \multirow{3}{*}{\shortstack{Unseen\\1 Drug}} & ACC  & 68.0 \textpm 0.7$^2$ & 68.6 \textpm 0.1 & 68.6 \textpm 0.2 & 67.8 \textpm 0.2 & 67.2 \textpm 0.2\\
                                        & AUROC & 88.6 \textpm 0.7$^3$ & 86.3 \textpm 0.3 & 86.0 \textpm 0.2 & 85.8 \textpm 0.2 & 85.7 \textpm 0.1\\
                                        & AUPR  & 73.7 \textpm 1.1$^3$ & 73.1 \textpm 0.3 & 72.1 \textpm 0.2 & 71.4 \textpm 0.5 & 71.6 \textpm 0.3\\
        \midrule
        \multirow{3}{*}{\shortstack{Unseen\\2 Drugs}} & ACC  & 53.6 \textpm 2.2$^2$ & 52.9 \textpm 0.6 & 52.9 \textpm 0.5 & 51.0 \textpm 0.6 & 50.1 \textpm 0.7\\
                                         & AUROC & 78.9 \textpm 1.5$^2$ & 70.9 \textpm 0.7 & 72.4 \textpm 0.4 & 70.9 \textpm 0.5 & 70.1 \textpm 0.6\\
                                         & AUPR  & 53.7 \textpm 1.9$^2$ & 51.5 \textpm 0.5 & 51.8 \textpm 0.4 & 49.9 \textpm 0.6 & 49.4 \textpm 0.6\\
        \bottomrule
    \end{tabular}
    \vspace{0.5em}
    \begin{minipage}{0.95\linewidth}
        \small \textit{Note:} $^1$Results from CGIB \citep{cgib}. $^2$Results from MeTDDI \citep{motifsddi}. $^3$Results from Molormer \citep{molormer}.
    \end{minipage}
\end{table}

\subsection{Benchmark vs Knowledge Graph-Based Models}

\begin{table}[ht]
    \centering
    \caption{Ablation Study: Performance of Different Embeddings in DDI Prediction of the DrugBank Dataset Proposed by \citep{knowddi}.)}
    \vspace{0.15in}
    \label{tab:knowddi_ablation_results}
    \begin{tabular}{@{}lccccc@{}}
        \toprule
        \textbf{Metric} & \textbf{KnowDDI} & \textbf{MFPS} & \textbf{GCN\_T} & \textbf{GCN} & \textbf{MoLF} \\
        \midrule
        F1 & 91.53 $\pm$ 0.24 & \textbf{95.38 $\pm$ 0.06} & 93.02 $\pm$ 0.08 & 92.44 $\pm$ 0.06 & 95.17 $\pm$ 0.07 \\
        ACC & 93.17 $\pm$ 0.09 & \textbf{95.41 $\pm$ 0.06} & 93.07 $\pm$ 0.08 & 92.49 $\pm$ 0.06 & 95.21 $\pm$ 0.07 \\
        COHEN & 91.89 $\pm$ 0.11 & \textbf{94.50 $\pm$ 0.08} & 91.69 $\pm$ 0.09 & 91.00 $\pm$ 0.07 & 94.26 $\pm$ 0.08 \\
        \bottomrule
    \end{tabular}
\end{table}

\subsection{DTA Benchmark Results}\label{dta_results}

\begin{table}[ht]
    \centering
    \caption{Ablation Study: Performance on Affinity Prediction in the FDA Dataset Proposed by \citep{motifsddi}.}
    \vspace{0.15in}
    \label{tab:dta_ablation_results}
    \begin{tabular}{@{}lcc@{}}
        \toprule
        \textbf{Embedding} & \textbf{RMSE} & \textbf{PCC} \\
        \midrule
        MeTDDI   & 0.912  & 0.725 \\
        GCN    & 0.983 $\pm$ 0.028 & 0.717 $\pm$ 0.019 \\
        GCN\_T & 1.016 $\pm$ 0.015 & 0.6928 $\pm$ 0.0115\\
        MFPS   & \textbf{0.904 $\pm$ 0.041}  & \textbf{0.782 $\pm$ 0.033} \\
        MOLF   & 1.242 $\pm$ 0.044 & 0.400 $\pm$ 0.121\\
        \bottomrule
    \end{tabular}
\end{table}

\newpage

\section{Additional Motif Analysis}\label{motif_add}

This section presents a comprehensive overview of the most common motifs found in each embedding method. The table below includes the top 3 motifs foun by MFPS and sulfur-containing motifs identified.

\begin{table}[ht]
    \centering
    \caption{Top 3 Motifs and Sulfur-Containing Motifs Across Embeddings.}
    \vspace{0.15in}
    \label{tab:full_motif_comparison}
    \begin{tabular}{@{}lcc|cc|cc@{}}
        \toprule
         & \multicolumn{2}{c}{\textbf{MFPS}} & \multicolumn{2}{c}{\textbf{MoLFormer}} & \multicolumn{2}{c}{\textbf{GCN-Trained}} \\
        \cmidrule(lr){2-3} \cmidrule(lr){4-5} \cmidrule(lr){6-7}
         \textbf{Motif} & \textbf{Ritonavir} & \textbf{Cobicistat} & \textbf{Ritonavir} & \textbf{Cobicistat} & \textbf{Ritonavir} & \textbf{Cobicistat} \\
        \midrule
        \textbf{Top 3 Motifs} & & & & & & \\
        CC(C)NC(N)=O & 322 & 163 & - & - & - & - \\
        CN(C)C & 308 & 129 & 63 & 118 & 11 & 58 \\
        ccsc(C)n (Thiazole) & 259 & 292 & - & - & - & - \\
        \midrule
        \textbf{Sulfur Motifs} & & & & & & \\
        ccsc(C)n (Thiazole) & 259 & 292 & - & - & - & - \\
        cc(C)nc(C)s (Thiazole) & 206 & 276 & - & - & - & - \\
        cnc(cs)CN (Sulfur-nitrogen) & 110 & 100 & - & - & - & - \\
        Cc(n)s (Thiazole variant) & 1 & - & 27 & 157 & 21 & 31 \\
        \bottomrule
    \end{tabular}
\end{table}

\begin{figure}[h!]
    \centering
    \includegraphics[width=0.48\textwidth]{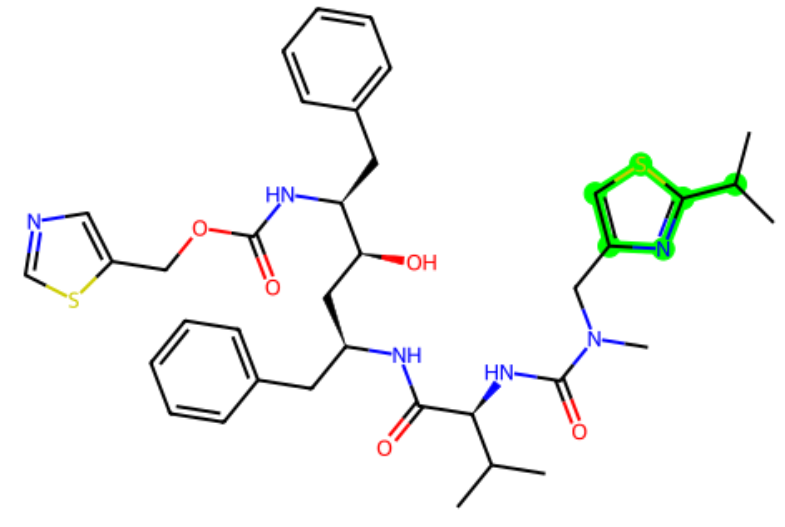}
    \hfill
    \includegraphics[width=0.48\textwidth]{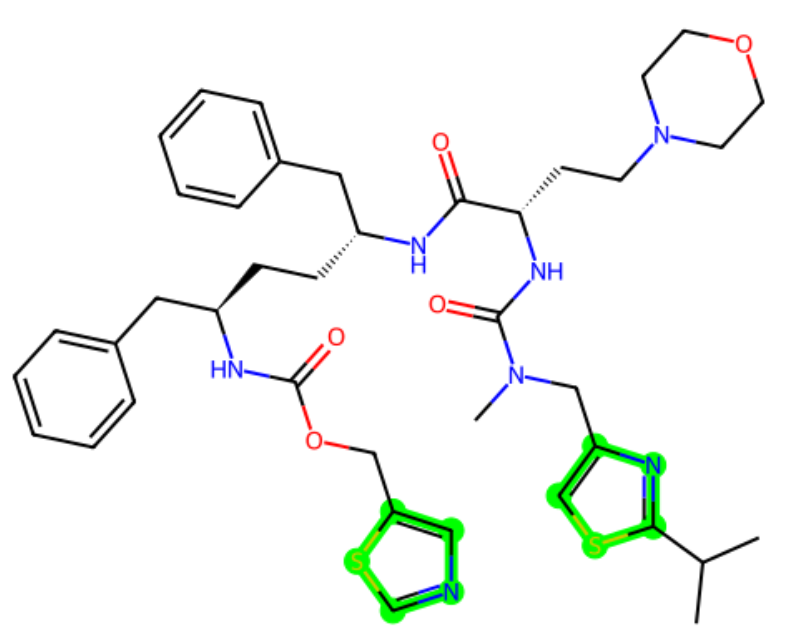}
    \caption{Highlighted sulfur-containing motifs in Ritonavir (left) and Cobicistat (right) found using the MFPS encodings.}
    \label{fig:highlighted_motifs}
\end{figure}

\subsection{Justification and Observations}

The motif analysis reveals distinct trends across embeddings, aligning with known molecular interactions:

\textbf{MFPS Embeddings:}  
MFPS effectively capture thiazole cores and sulfur-nitrogen motifs, both critical for CYP3A4 inhibition, reinforcing their suitability for modeling drug interactions. The presence of multiple thiazole-based motifs suggests strong aromatic stabilization in ligand binding.

\textbf{MoLFormer Embeddings:}  
MoLFormer primarily identifies CN(C)C motifs associated with hydrogen bonding and polar interactions. It also captures a thiazole variant (Cc(n)s) in both molecules, indicating a different weighting of molecular substructures compared to fingerprint-based methods.

\textbf{Pretrained GCN Embeddings:}  
Pretrained GCN embeddings detect fewer motifs overall but capture a subset similar to those identified by MoLFormer, including thiazole-related structures.

\newpage

\section{Visualization of DDA Prediction Performance}\label{dda_mfps_plot}

\begin{figure}[ht]
    \centering
    \includegraphics[width=\linewidth]{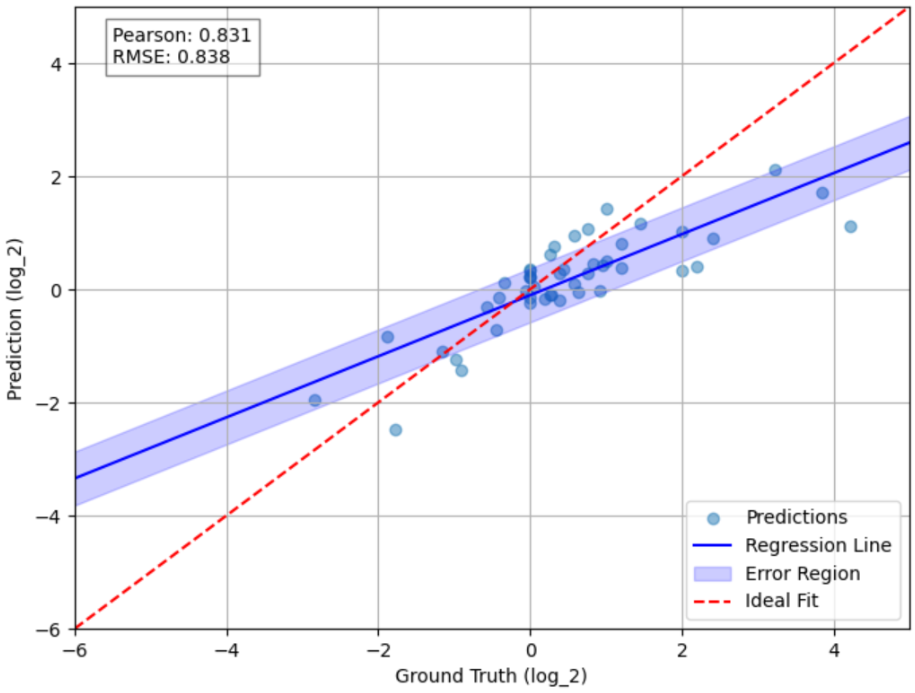}
    \caption{Performance visualization of the DDA regression model using MFPS embeddings. The results highlight the relationship between predicted and ground truth values of $\log_2(\text{AUC FC})$.}
    \label{fig:dda_mfps}
\end{figure}

\end{document}